\documentclass{aa}
\usepackage{times}
\usepackage{graphics}
\newcommand\jcd{Christensen-Dalsgaard}

\newcommand\phn{\ensuremath{\phantom{0}}}

\newcommand\ea{et al.}
\newcommand\aap{A\&A}

\newcommand\apj{ApJ}

\newcommand\mnras{MNRAS}

\newlength{\figwidth}
\begin{document}
\thesaurus{08.09.3; 08.15.1; 08.18.1}
\title{Seismic detection of stellar tachoclines}
\titlerunning{Seismic detection of stellar tachoclines}
\author{Anwesh Mazumdar \and H. M. Antia}
\authorrunning{Mazumdar \and Antia}
\offprints{H. M. Antia}
\institute{Tata Institute of Fundamental Research,
Homi Bhabha Road, Mumbai 400005, India\\
email: anwesh@tifr.res.in, antia@tifr.res.in}
\date{Received }
\maketitle

\begin{abstract}
Helioseismic inversions for the rotation rate have established the
presence of a tachocline near the base of the solar convection zone.
We show that the tachocline produces a characteristic
oscillatory signature in the splitting coefficients of low degree
modes, which could be observed on distant stars. Using this
signature it may be possible to determine the characteristics of
the tachocline using only low degree modes. The limitations of
this technique in terms of observational uncertainties are
discussed, to assess the possibility of detecting tachoclines on
distant stars.

\end{abstract}

\section{Introduction}
\label{sec:intro}

Helioseismic inversions of observed splitting coefficients have
enabled us to study the rotation rate inside the Sun (Thompson
\ea\ \cite{thompson:96}; Schou \ea\ \cite{schou:98}). 
From these inversions it has
been established that there is a shear layer near the base of the
convection zone where the rotation rate undergoes a
transition from differential rotation inside the
convection zone to almost uniform rotation in the radiative
interior. This layer has been referred to as the
tachocline (Spiegel \& Zahn \cite{sz:92}).
The characteristics of this layer have been studied using
helioseismic data (Kosovichev \cite{koso:96}; Basu \cite{basu:97}; 
Antia, Basu \& Chitre \cite{abc:98}; Charbonneau \ea\ \cite{charbon:99};
Corbard \ea\ \cite{corbard:99}).
Nevertheless, the origin of this shear layer is not yet clear and it 
would be instructive to probe the possible existence of these
layers in distant stars. Such a study would help us in understanding
the formation of tachoclines and to test the theories of angular
momentum transport in stellar interiors.

The solar tachocline has been detected using frequency splittings
for modes of low and intermediate degree, $\ell$. All these modes
are not expected to be detected on other stars. In order to
detect a tachocline on distant stars we have to look for the signature
of a tachocline in low degree modes ($\ell\le3$), which are the
only modes that can be detected on these stars. It has been
shown that rapid variations in the sound speed in the stellar
interior, like those arising at the base of the convection zone
leave a characteristic oscillatory signature in the mean frequencies
of low degree modes (Gough \cite{gough:90}; Monteiro, \jcd\ \& Thompson
\cite{mct:94}; Roxburgh \& Vorontsov \cite{rv:94}; 
Basu, Antia \& Narasimha \cite{ban:94}). 
From this oscillatory signature
it has been possible to put limits on the extent of overshoot
below the solar convection zone (Basu \cite{basu:97}). Monteiro, \jcd\
\& Thompson (\cite{mct:00}) have pointed out that this oscillatory signature 
can be used to study the location of the base of the convection zone
as well as the extent of overshoot below this base in other stars
using asteroseismic data for only low degree modes.

In this work we study the signature of a tachocline in the low degree
modes. Since the tachocline is a narrow layer where the rotation rate
varies rapidly, we would expect an oscillatory signature in the
corresponding splitting coefficients. Using a model for the solar
tachocline, we show that this oscillatory signature is indeed
present in frequency splitting coefficients of the low degree modes.
Further, it is, in principle, possible to determine the characteristics
of the tachocline, like its location, width and the extent of variation
in the rotation rate across the tachocline using this oscillatory
signature.

\section{The technique}
\label{sec:technique}

The frequency
of an eigenmode of a given degree $\ell$, radial order, $n$, and
azimuthal order, $m$ can be expressed in
terms of the splitting coefficients, using the expansion
\begin{equation}
\nu_{n\ell m} = {\omega_{n\ell m}\over 2\pi}
= \nu_{n\ell} + \sum_{j=1}^{j_{\rm max}} c_j (n,\ell) \,
{\cal P}_j^{(\ell)}(m).
\label{eq:split}
\end{equation}
Here $\nu_{n\ell}$ is the mean frequency of the $(n,\ell)$ multiplet,
$c_j(n,\ell)$ are the splitting coefficients and
${\cal P}_j^{(\ell)}(m)$ are orthogonal polynomials in $m$
(Ritzwoller \& Lavely \cite{rl:91}). If the rotation energy in the star is
much smaller than the gravitational energy, then rotation can be
treated as a small perturbation over the non-rotating spherically
symmetric star. In this approximation the splitting coefficients
can be determined by the corresponding component of the rotation rate.
Following Ritzwoller \& Lavely (\cite{rl:91}), we express the rotation
rate as
\begin{equation}
\Omega(r,\theta)=
-\sum_j {w_{2j+1}(r)\over r\sin\theta}
{d\over d\theta}P_{2j+1}(\cos\theta),
\label{eq:rot}
\end{equation}
where $\theta$ is the colatitude, $P_j$ are the Legendre polynomials
and $w_j(r)$ are expansion coefficients for the
rotation rate which are related to the splitting coefficients
$c_j(n,\ell)$ by
\begin{equation}
c_j(n,\ell)=\int_0^R w_j(r)K_j^{(n,\ell)}(r)r^2\;dr,
\label{eq:split_def}
\end{equation}
where the kernels $K_j^{(n,\ell)}(r)$ are defined in terms of
eigenfunctions of the mode (Ritzwoller \& Lavely \cite{rl:91}).
For the Sun it is found that a major part of the transition in the
tachocline is determined by the splitting coefficient $c_3(n,\ell)$
(Antia, Basu \& Chitre \cite{abc:98}). The corresponding component of
the rotation rate is defined by
\begin{equation}
\Omega_3(r,\theta)=-{w_3(r)\over r\sin\theta}{d\over
d\theta}P_3(\cos\theta)={w_3(r)\over r}{3\over 2}
(5\cos^2\theta-1).
\label{eq:omega_3}
\end{equation}
The variation in the other components across the tachocline turns out to be
an order of magnitude smaller. The splitting coefficient $c_3(n,\ell)$
is defined only for $\ell\ge2$ and hence only these modes can
be used to study this component.
In the absence of any theoretical
understanding of formation of a tachocline, it is difficult to
predict the expected form or magnitude of variation for other stars.
In this work, we assume that in other stars too the dominant
variation is in the $\Omega_3$ component. Although,
the analysis is similar for all splitting coefficients,
the higher order coefficients can only be determined for modes of
higher degree. Since for other stars we can only expect to detect
modes with degree $l=0,1,2,3$, it is not possible to study
transition in higher order terms. On the other hand, if the
dominant variation in other stars is found in the spherically
symmetric component corresponding to $c_1(n,\ell)$ it will be
easier to detect the tachocline, but we will not consider this
possibility in this work.
 
In the asymptotic limit (\jcd\ \& Berthomieu \cite{cb:91})
we can approximate the kernel as
$~\cos^2(\omega\tau+\phi)$, where $\tau$ is the acoustic depth given by
\begin{equation}
\tau=\int_r^R {1 \over c(r)}\; dr.
\label{eq:tau}
\end{equation}
From  Eq.~\ref{eq:split_def} it is clear that if we have a rotation rate 
which is piecewise constant with transition at $r=r_d$, then the
corresponding splitting coefficient will have an oscillatory
term of the form $\sin(2\omega\tau+\phi)$, similar to that
observed in the mean frequency due to transition in the sound speed
(Monteiro \ea\ \cite{mct:94}; Basu \ea\ \cite{ban:94}). The oscillatory
part arises because the integral over the region covering an
entire wavelength of oscillation will give an average value,
while the last part between the nearest node of the eigenfunction and the
location of transition, $r_d$ will give an oscillatory contribution.
Thus we can express the splitting coefficient as
\begin{equation}
c_j(n,\ell)=c_j^{(s)}(n,\ell)+A(\omega)\sin\!\left(2\omega\tau+\phi-
{\gamma \ell(\ell+1)\over \omega}\right),
\label{eq:split_form}
\end{equation}
where $c_j^{(s)}(n,\ell)$ is the smooth part of the coefficient, which
may arise from possible smooth variations in the rotation rate with
depth. The amplitude $A(\omega)$ is a smooth function of $\omega$,
while $\gamma$ is a constant. The term involving $\gamma$ 
arises from a more accurate expression for
the vertical wavenumber $k_r$ (Monteiro \ea~\cite{mct:94}).
This term can be
neglected while considering only low degree modes, since it is
very small for low $\ell$. In general, the amplitude also may depend
on $\ell$, but at low degrees these terms are likely to be small.

Following Basu \ea~(\cite{ban:94}), we take the fourth difference of
the splitting coefficients with respect to $n$
to enhance the oscillatory signal. Another advantage of taking the
fourth difference is that the smooth part of the splitting coefficients
becomes negligible and we do not need to include it in our analysis.
This will of course, depend on the smooth component of variation
of the rotation rate, but even for a realistic solar rotation profile this
component is found to be negligible.

\begin{figure*}
\centering \leavevmode
\hbox to \hsize{\hfil
\resizebox{\figwidth}{!}{\includegraphics{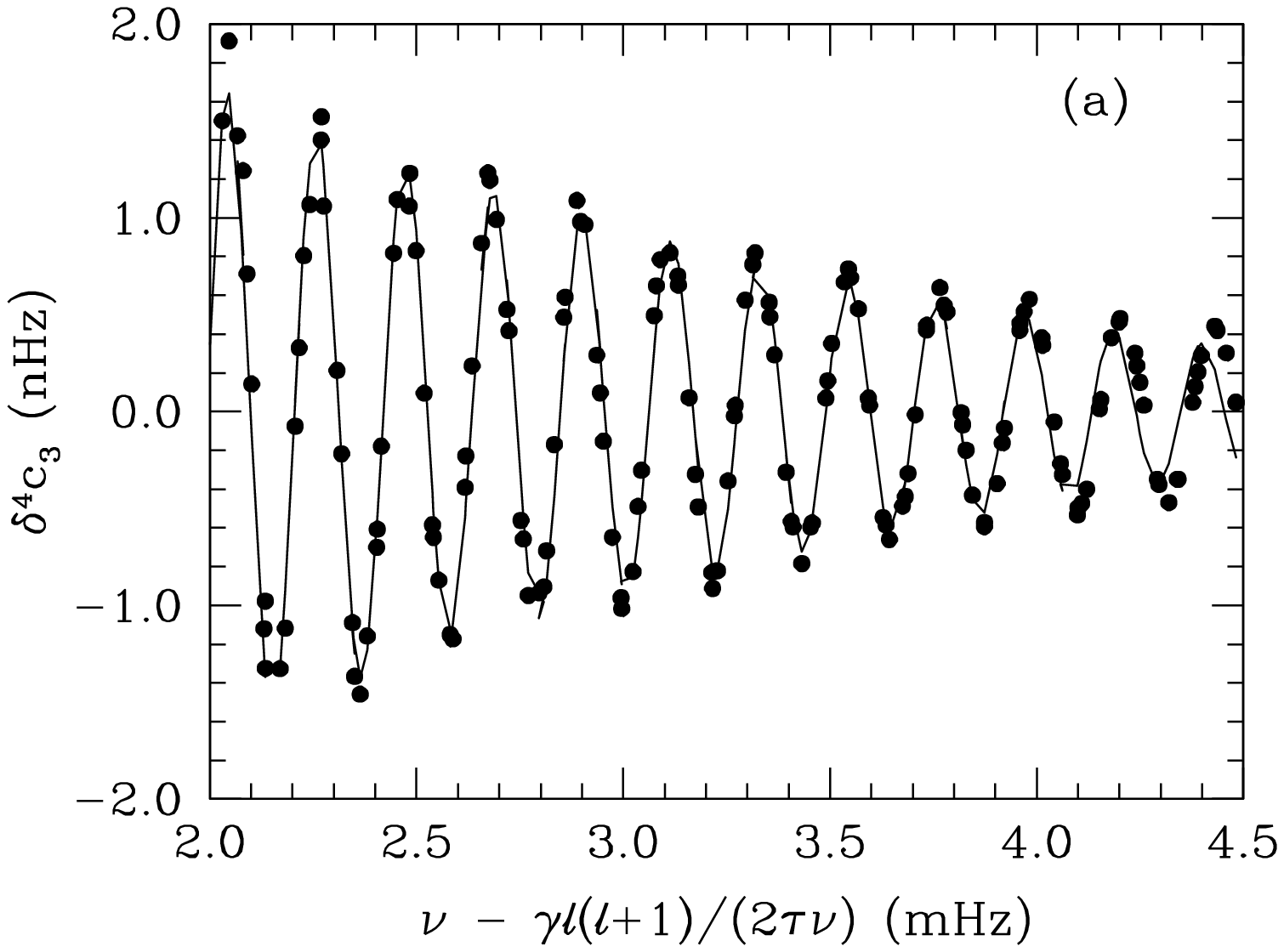}}
\hfil
\resizebox{\figwidth}{!}{\includegraphics{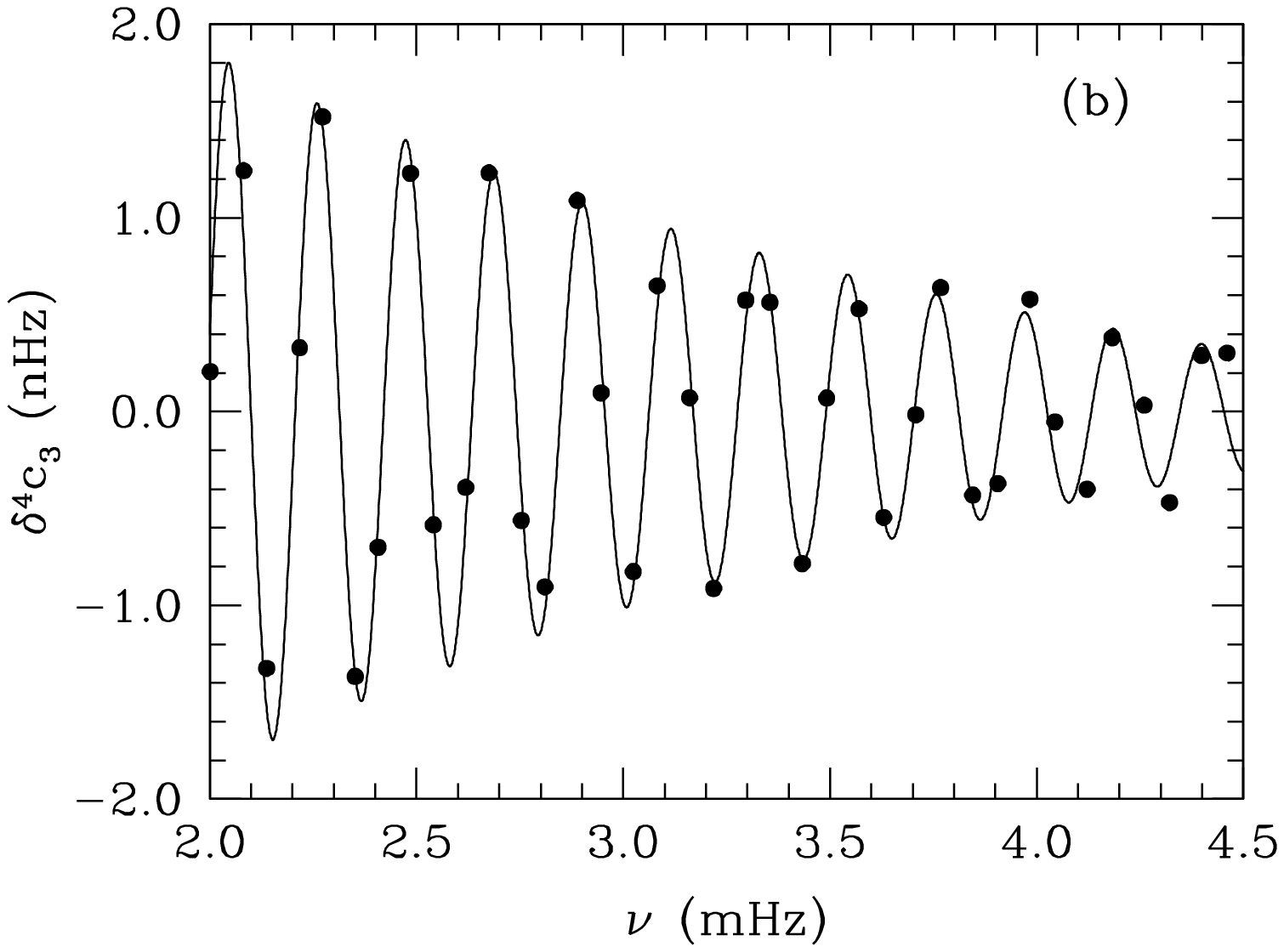}}
\hfil}
\caption{Fourth difference of the splitting
coefficient $c_3$, for $\delta\Omega=20$~nHz, $w=0.005R_\odot$
and $r_d=0.7R_\odot$.
The {\it continuous line} shows a fit according to Eq.~\ref{eq:fourth_fit}.
(a) Modes with $2 \le \ell \le 10$ have been included. 
The frequency has been shifted to account for the
$\ell$-dependence of the term in sine function.
(b) Only modes with $\ell =2$ and $\ell = 3$ have been included. 
	\label{fig:osc_fit}
}
\end{figure*}

In order to illustrate this oscillatory signature we assume a model
tachocline rotation profile of the form
\begin{equation}
\Omega(r,\theta)={\delta\Omega (5\cos^2\theta -1)\over
1+\exp[(r_d-r)/w]},
\label{eq:rotprof}
\end{equation}
where $\delta\Omega$ is the extent of variation in the rotation rate
across the tachocline, $w$ is the half-width of the transition
layer and $r_d$ is its mid-point.
It may be noted that this form is different from that used by
Kosovichev~(\cite{koso:96}) and Charbonneau \ea~(\cite{charbon:99}), 
and in particular,
as explained by Antia \ea~(\cite{abc:98}), the definition of
width is also different. The half-width $w$ in our definition
should be multiplied by $4.9$ to obtain the width as defined by
them.
Using this model rotation profile we can calculate the corresponding
splitting coefficients for the Sun. Because of the choice of
latitudinal dependence only the splitting coefficient $c_3(n,\ell)$
is found to be non-zero and Fig.~\ref{fig:osc_fit}(a)
shows the fourth difference of
these coefficients. For illustration we have included modes
with $2\le\ell\le 10$. The oscillatory trend is clearly seen in
these differences. The `wavelength' of these oscillations depends
on the depth $r_d$, while the amplitude is determined by the width
and $\delta\Omega$. Since the splitting coefficients and hence
the fourth differences scale linearly
with $\delta\Omega$ we have used only one value $\delta\Omega=20$~nHz
which is the typical variation across the solar tachocline.
In order to characterize the oscillatory signal
we fit the following form to these differences:
\begin{equation}
\delta^4 c_3(n,\ell)=\left(a_0+{a_1\over \omega}+{a_2\over\omega^2}\right)
\sin\!\left(2\omega\tau+\phi-{\gamma \ell(\ell+1)\over \omega}\right).
\label{eq:fourth_fit}
\end{equation}
The parameters $a_0,a_1,a_2,\tau,\phi,\gamma$
can be determined by a nonlinear least squares fit.

For distant stars it will be possible to detect modes with $\ell=0,1,2,3$
only and hence the number of modes will be highly restricted. Even then it is
possible to fit the oscillatory part and Fig.~\ref{fig:osc_fit}(b)
shows one such fit.
In this case the parameter $\gamma$ is not relevant as the corresponding
term is very small and we fit only the five parameters $a_0$, $a_1$, $a_2$,
$\tau$ and $\phi$.
All the results presented in this paper are obtained using modes
for $\ell = 2,3$ only, unless mentioned otherwise.

\section{Results}
\label{sec:results}

\begin{figure}
\begin{center}
\resizebox{\figwidth}{!}{\includegraphics{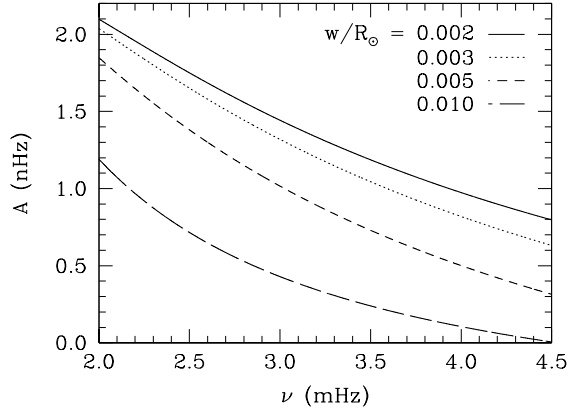}}
\end{center}
\caption{
Comparison of amplitude of oscillatory signal as a function of frequency
for different widths of the transition region for $\delta\Omega=20$~nHz
and $r_d=0.7R_\odot$.
\label{fig:amp_comp}
}
\end{figure}

In order to study the signature of the tachocline in the splitting coefficients
we calculate $c_3(n,\ell)$ with rotation profiles defined by
Eq.~\ref{eq:rotprof}
using different values of half-width $w$, and position $r_d$ of the
tachocline. We fit the oscillatory form defined by
Eq.~\ref{eq:fourth_fit} to each of these data sets.
It turns out that the position of the tachocline affects mainly the parameter
$\tau$, which is close to the acoustic depth of the transition layer.
Therefore, in this work we have only shown results with $r_d=0.7R_\odot$.
Conversely, the location of the tachocline may be determined
from the parameter $\tau$. In order to study the effect of width
we use different values of $w$ and the fitted amplitude
($A(\omega)=a_0+a_1/\omega+a_2/\omega^2$) is shown in
Fig.~\ref{fig:amp_comp}. It can be
seen that in all cases the amplitude decreases with increasing
$\omega$, which may be expected as the modes with higher frequency
have larger number of nodes in radial direction and hence the
radial wavelength will be smaller, thus giving a smaller contribution
to the integral in Eq.~\ref{eq:split_def}. 
At larger width the amplitude reduces
and the tachocline profile also gives a contribution to the smooth
part at low frequencies and it is not possible to use the simple
form to fit the data.
It turns out that the extent of
variation in the amplitude across the frequency range depends on the
width of the tachocline. For small width the variation is smaller, while
with increasing width, the variation in amplitude increases. For example,
for $w=0.002R_\odot$, $0.003R_\odot$, $0.005R_\odot$ and $0.01R_\odot$,
the ratios of amplitudes at the two limits in
Fig.~\ref{fig:amp_comp} are $2.5$, $3.0$, $5.5$ and $100$ respectively. 
Thus, if we have  data covering
the entire frequency range it should be possible to determine the
width using the extent of variation in amplitude. Once the width is
determined, we can determine $\delta\Omega$ from the amplitude of the
oscillatory term at the low frequency end. Thus we should be able to 
determine the characteristics like position, width and $\delta\Omega$ of the
tachocline using only low degree modes. 

In all the foregoing calculations we have used the exact splitting coefficients
as calculated for a given rotation profile. In real data, there would
naturally be errors associated with each splitting coefficient. In order to
simulate real data we have constructed artificial data sets where
random errors with standard deviation $\sigma$ are added to all the
splitting coefficients. For simplicity, we assume that error is same
in all these coefficients. Using $100$ different realizations of errors we
can estimate the expected errors in each of the fitted parameters, the
results of which are summarized in Table~\ref{tab:param}. For low errors
($\sigma\la 0.02$~nHz) it is
indeed possible to determine all the parameters to reasonable
accuracy and the error in each parameter is proportional to the assumed
error in the splitting coefficients.
\begin{table}
\centering
\caption{Values of fitted parameters in oscillatory signal for
$\delta\Omega=20$~nHz, $w=0.005R_\odot$, $r_d=0.7R_\odot$
\label{tab:param}
}
\begin{tabular}{cccc}
\hline
Error ($\sigma$) & $a_0$ & $\tau$ & $\phi$\\
(nHz) & (nHz) & (sec)  & (rad)\\
\hline
$0.010$ & $0.68 \pm 0.02$ & $2323 \pm 5\phn\phn$ & $4.70 \pm 0.14$ \\
$0.020$ & $0.68 \pm 0.04$ & $2324 \pm 10\phn   $ & $4.68 \pm 0.28$ \\
$0.050$ & $0.70 \pm 0.10$ & $2331 \pm 32\phn   $ & $4.59 \pm 0.81$ \\
$0.100$ & $0.81 \pm 0.19$ & $2330 \pm 135      $ & $4.53 \pm 1.40$ \\
$0.200$ & $1.25 \pm 0.45$ & $2290 \pm 242      $ & $4.79 \pm 1.63$ \\
$0.200$ & $0.79 \pm 0.36$ & $2323$ & $4.70 \pm 0.78$ \\
\hline
\end{tabular}
\end{table}

However, as error increases it is not possible to determine
the amplitude variation correctly, but if we fix $a_1=0$ and $a_2=0$
then we can still determine the mean amplitude of the oscillatory part.
For $\delta\Omega=20$~nHz, we find that it is possible to detect
the oscillatory signal reliably when the error $\sigma\la 0.2$~nHz, which
corresponds to an error of about $1.6$~nHz in the fourth differences.
If we assume that the tachocline in other stars would also be located
close to the base of the convection zone, then its depth can be
determined by looking at the oscillatory part in the mean frequencies
(Monteiro \ea~\cite{mct:00}) which determines the base of the
convection zone. In that case we can fix $\tau$ to the value
determined from the convection zone base to improve the detectability of
the oscillatory signal. The last line in Table~\ref{tab:param}
shows the results
of simulations obtained when $\tau$ was kept fixed. It is clear that
the estimated errors in this case are smaller.
If we have the splitting coefficients for only $\ell = 2$ modes then the
errors are found to be $\sqrt{2}$ times larger.
For the solar case we can use higher degree modes also, but even in
that case the error in the observed splitting coefficients is slightly
larger than this limit and it is not possible to detect
the signal reliably.

\section{Discussion}
\label{sec:discuss}

We have shown that a sharp transition in the rotation rate that is
expected in the tachocline region gives rise to an oscillatory signal in
the splitting coefficients of the low degree modes. The
`wavelength' of the oscillatory signal is determined by the position
of the tachocline, while the variation in amplitude across the frequency
range is determined by its width. The amplitude of signal is of course
proportional to the extent of the variation in the rotation rate. Thus from
this oscillatory signal it is, in principle,
possible to determine the position,
width and $\delta\Omega$ for the tachocline. 
We are looking for the signal in odd splitting coefficients, which can
arise only due to rotation, and not
from magnetic fields or structural variations in the stellar interior.
In this discussion
we have not included any possible latitudinal variation in the
characteristics of the tachocline. The solar tachocline is known
to be prolate (Charbonneau \ea~\cite{charbon:99}) and this variation would
also affect the oscillatory signal in low degree modes. Using only
low degree modes it may not be possible to disentangle all variations
in the tachocline, but so long as the latitudinal variation is small, as
is the case for the Sun, the mean properties of the tachocline can
be determined from the low degree modes.

Since the amplitude of the oscillatory signal is very small, it
will be necessary to find the splitting coefficients accurately
to determine the characteristics of the tachocline. From our
simulations it appears that the required accuracy in the splitting
coefficients is $\sim 0.2$~nHz for
$\delta\Omega=20$~nHz. Even for the Sun, this level of accuracy
has not been achieved so far and we do not expect it to be achieved for
other stars in near future.
But there is no reason to assume that the variation in rotation
rate in all stars will be comparable to that in the Sun.
In particular, for stars which are fast rotators, we would expect
$\delta\Omega$ to be correspondingly larger.
Even for stars with similar rotation rate there may be some
variation in $\delta\Omega$ or in the amplitude of oscillatory signal
for the same $\delta\Omega$. In this work, all the calculations have been
done for a solar model. In other stars the amplitude may be
somewhat different.  If the
star is rotating very rapidly, the linear approximation used to
relate the splitting coefficients to the rotation rate may not
be admissible. But we would expect that for a star which is rotating
about $50$ times faster than the Sun, this approximation may still
be applicable and in that case an accuracy of about $10$~nHz may be sufficient
to detect the oscillatory signal.
For stars with larger $M/R^3$, this limit may also be larger.
Similarly, if we choose stars with larger differential rotation, or with
favourable amplitude, this number may go up further.
Even if stars are rotating more rapidly the oscillatory signal may
still be present, but mode identification and interpretation may
be more difficult.
The upcoming asteroseismic missions like COROT (Baglin \ea\
\cite{baglin:98}), MOST (Matthews \cite{matthews:98}) and MONS 
(Kjeldsen \& Bedding \cite{kb:98})
have a planned frequency resolution of $100$~nHz, which may not be
sufficient to detect this oscillatory signal.
But with some improvements in instruments and longer observations of
a few selected stars, which are known to be rotating fast
and preferably having larger differential rotation, it may be
possible to detect this signal in not too distant future.

\end{document}